\documentclass[dvips]{arxstspdf}
\usepackage{dcolumn}
\usepackage{graphicx}
\usepackage{flushend}
\usepackage{stfloats}

\volume{23}
\issue{3}
\pubyear{2008}
\firstpage{404}
\lastpage{419}
\doi{10.1214/08-STS269}

\makeatletter
\newcolumntype{d}[1]{D{.}{.}{#1}}
\makeatother

\begin{document}
\begin{frontmatter}

\title{Handling Covariates in the Design of Clinical Trials}
\runtitle{Covariates in Clinical Trials}

\begin{aug}
\author[a]{\fnms{William F.} \snm{Rosenberger}\corref{}\ead[label=e1]{wrosenbe@gmu.edu}} \and
\author[b]{\fnms{Oleksandr} \snm{Sverdlov}\ead[label=e2]{alex.sverdlov@bms.com}}
\runauthor{W. F. Rosenberger and O. Sverdlov}

\affiliation{George Mason University and Bristol-Myers Squibb}

\address[a]{William F. Rosenberger is Professor and Chair,
Department of Statistics, George Mason University,
4400 University Drive, MS 4A7 Fairfax, Virginia
22030-4444, USA, \printead{e1}.}
\address[b]{Oleksandr Sverdlov is Senior Research Biostatistician,
Bristol-Myers Squibb, Route 206 and Province Line Road,
Lawrenceville, New Jersey 08540, USA, \printead{e2}.}

\end{aug}

% ABSTRACT
%
\begin{abstract}
There has been a split in the statistics community about the need
for taking covariates into account in the design phase of a clinical
trial. There are many advocates of using stratification and
covariate-adaptive randomization to promote balance on certain known
covariates. However, balance does not always promote efficiency or
ensure more patients are assigned to the better treatment. We
describe these procedures, including model-based procedures, for
incorporating covariates into the design of clinical trials, and
give examples where balance, efficiency and ethical considerations
may be in conflict. We advocate a new class of procedures,
covariate-adjusted response-adaptive (CARA) randomization
procedures that attempt to optimize both efficiency
and ethical considerations, while maintaining randomization. We
review all these procedures, present a few new simulation studies,
and conclude with our philosophy.
\end{abstract}

% KEYWORDS
%
\pdfkeywords{Balance, covariate-adaptive randomization,
covariate-adjusted response-adaptive randomization, efficiency, ethics}

\end{frontmatter}

%s1 ###
\section{Introduction}\label{s1}

Clinical trials are often considered the ``gold standard'' in
convincing the medical community that a therapy is beneficial in
practice. However, not all clinical trials have been universally
convincing. Trials that have inadequate power, or incorrect
assumptions made in planning for power, imbalances on important
baseline covariates directly related to patient outcomes, or
heterogeneity in the patient population, have contributed to a lack
of scientific consensus. Hence, it is generally recognized that the
planning and design stage of the clinical trial is of great
importance. While the implementation of the clinical trial can often
take years, incorrect assumptions and forgotten factors in the
sometimes rushed design phase can cause controversy following a trial.
For example, take the trial of
erythropoietin in maintaining normal hemoglobin concentrations in
patients with metastatic breast cancer (Leyland-Jones, \citeyear{ley2003}). This
massive scientific effort involved 139 clinical sites and 939
patients. The study was terminated early because of an increase in
mortality in the erythropoietin group. The principal investigator\break
explains:

\begin{quote}
\ldots drawing definitive conclusions has been
difficult because the study was not designed to prospectively
collect data on many potential prognostic survival factors that
might have affected the study outcome$\ldots.$ The results of this trial
must be interpreted with caution in light of the potential for an
imbalance of risk factors between treatment groups$\ldots.$ The
randomisation design of the study may not have fully protected
against imbalances because the stratification was only done for one
parameter$,\ldots$ and was not done at each participating centre$\ldots.$ It
is extremely unfortunate that problems in design\ldots\ have complicated
the interpretation of this study. Given the number of design issues
uncovered in the post hoc analysis, the results cannot be considered
conclusive.
\end{quote}
An accompanying commentary calls this article\break
``alarmist,'' thus illustrating the scientific conundrum that
covariates present in clinical trials. There is no agreement in the
statistical community about how to deal with potentially important
baseline covariates in the design phase of the trial. Traditionally,
prestratification has been used on a small number of very important
covariates, followed by stratified analyses. But what if the
investigator feels there are many covariates that are important---too
many, in fact, to feasibly use prestratification?

The very act of randomization tends to mitigate the probability that
important covariates will be distributed differently among treatment
groups. This property is what distinguishes randomized clinical
trials from observational studies. However, this is a large sample
property, and every clinical trialist is aware of randomized trials
that resulted in significant baseline covariate imbalances. %\textbf{
Grizzle (\citeyear{griz1982}) distinguished two factions of the statistical
community, the ``splitters'' and the ``lumpers.'' The splitters
recommend incorporating important covariates into randomization,
thus ensuring balance over these covariates at the design stage. The
lumpers suggest ignoring covariates in the design and use simple
randomization to allocate subjects to different treatment groups,
and adjust for covariates at the analysis stage. %The adjustment for
%unbalanced covariates is one of the key points of the clinical trial
%practice.
%}
As Nathan Mantel once pointed out (Gail, \citeyear{gai1992}):

\begin{quote}
\ldots After looking at a data set, I might see
that in one group there are an unusually large number of males. I
would point out to the investigators that even though they had
randomized the individuals to treatments, or claimed that they had,
I could still see that there was something unbalanced. And the
response I would get was ``Well, we randomized and therefore we
don't have to bother about it.'' But that isn't true. So, as long
as the imbalance is an important factor you should take it into
account. Even though it is a designed experiment, in working with
humans, you cannot count on just the fact that you randomized.
\end{quote}
Today, many statisticians would argue that the only
legitimate adjusted analyses are for prespecified important
covariates planned for in the analysis according to protocol, and
that these adjustments should be done whether or not the
distributions are imbalanced (e.g., Permutt, \citeyear{per2000}). In addition,
these covariates should be accounted for in the \textit{design} of the
trial, usually by prestratification, if possible.

The three-stage philosophy of prestratifying on important known
covariates, followed by a stratified analysis, and allowing for
randomization to ``take care of'' the other less important (or
unknown) covariates, has become a general standard in clinical
trials. This method breaks down, however, when there are a large
number of important covariates. This has led to the introduction of
\textit{covariate-adaptive randomization} procedures, sometimes
referred to as \textit{minimization} procedures or \textit{dynamic
allocation}.%1
\footnote{Or sometimes, unfortunately, as
just \textit{adaptive designs}, which could refer to any number of
statistical methods having nothing to do with covariates, including
response-adaptive randomization, sequential monitoring, and
flexible interim decisions.}
Some of these ``covariate-adaptive''
procedures (the term we will use) that have been proposed have been
randomized, and others not.

There is no consensus in either the statistics world or the clinical
trials world as to whether and when these covariate-adaptive
procedures should be used, although they are gaining in popularity
and are now used frequently. Recently clinical trialists using
these procedures have grown concerned that regulatory agencies have
expressed skepticism and caution about the use of these techniques.
In Europe, The Committee on Proprietary Medicinal Products (CPMP)
Points to Consider Document (see Grouin, Day and Lewis, \citeyear{groDayLew2004})
states:

\begin{quote}
Dynamic allocation is strongly
discourag\-ed$\ldots.$ Without adequate and appropriate
supporting/sensitivity analysis, an application is unlikely to be
successful.
\end{quote}
This document has led to much controversy. In a
commentary, Buyse and McEntegart (\citeyear{buyMce2004}) state:

\begin{quote}
In our view, the CPMP's position is unfair,
unfounded, and unwise$\ldots.$ It favors the use of randomization
methods that expose trialists and the medical community to the risk
of accidental bias, when the risk could have been limited through
the use of balancing methods that are especially valuable$\ldots.$ If
there were any controversy over the use of minimization, it would be
expected of an independent agency to weigh all the scientific
arguments, for and against minimization, before castigating the use
of a method that has long been adopted in the clinical community.
\end{quote}
In a letter to the editor, Day, Grouin and Lewis (\citeyear{dayGrLew2005})
respond that

\begin{quote}
\ldots the scientific community is not of one
mind regarding the use of covariate-adaptive randomization
procedures$\ldots.$ Rosenberger and Lachin cautiously state that ``very
little is known about its theoretical properties.'' This is a
substantial point. The direct theoretical link between randomization
and methods of statistical analysis has provided a solid foundation
for reliable conclusions from clinical trial work for many years.
\end{quote}

It is in the context of this controversy that this paper is written.
The intention of this paper is to explore the role of covariates in
the \textit{design} of clinical trials, and to examine the burgeoning
folklore in this area among practicing clinical trialists. Just
because a technique is widely used does not mean that it is
valuable. And just because there is little theoretical evidence
validating a method does not mean it is not valid. The
nonspecificity of the language in these opinion pieces is becoming
troubling: what is meant by the terms ``minimization,'' ``dynamic,''
``adaptive''? Many procedures to mitigate covariate imbalances have
been proposed. Are they all equally effective or equally
inappropriate? We add to the controversy by discussing the often
competing criteria of balance, efficiency and ethical
considerations. We demonstrate by example that clinical trials that
balance on known covariates may not always lead to the most
efficient or the most ethically attractive design, and vice versa.

This paper serves as both a review and a summary of some of our
thoughts on the matter; in particular, we advocate a new class of
procedures called \textit{covariate-adjusted response-adaptive} (CARA)
randomization procedures
(e.g., Hu and Rosenberger, \citeyear{huRos2006}).
The outline of the paper is as follows. In Section~\ref{s2}, we review the
most popular covariate-adaptive randomization procedures. In
Section~\ref{s3}, we describe randomization-based inference and its
relationship to clinical trials employing covariate-adaptive
randomization methods. In Section~\ref{s4}, we discuss what is known from
the literature about the properties of the procedures in Section~\ref{s2}.
In Section~\ref{s5}, we describe the alternative model-based optimal design
approach to the problem and describe properties of these procedures
in Section~\ref{s6}. In Section~\ref{s7}, we discuss the relationship between
balance, efficiency and ethics, and describe philosophical arguments
about whether balance or efficiency is a more important criterion.
We demonstrate by example that balance does not necessarily imply
efficiency and vice versa, and demonstrate that balanced and
efficient designs do not necessarily place more patients on the
better treatment. In Section~\ref{s8}, we describe CARA randomization
procedures and their properties. In Section~\ref{s9}, we report the results
of a simulation study comparing different CARA and
covariate-adaptive randomization procedures for a binary response
trial with covariates. %We show that CARA randomization procedures
%can be a good alternative to balanced designs in the nonlinear
%response case.
%clinical trials using heteroscedastic and nonlinear models.
Finally, we %discuss all these
%procedures and
give a summary of our own opinions in Section~\ref{s10}.

%s2 ###
\section{Covariate-Adaptive Randomization}\label{s2}

Following Rosenberger and Lachin (\citeyear{rosLac2002}), a \textit{randomization
sequence} for a two-treatment clinical trial of $n$ patients is a
random vector $\mathbf{T}_n=(T_1,\ldots,T_n)^{\prime}$, where $T_j=1$ if
the $j$th patient is assigned to treatment $1$ and $T_j=-1$ if the
patient is assigned to treatment $2$. A \textit{restricted
randomization procedure} is given by
$\phi_{j+1}=\Pr(T_{j+1}=1|\mathbf{T}_j)$, that is, the probability that the
$(j+1)$th patient is assigned to treatment 1, given the previous
$j$ assignments. When the randomization sequence is dependent on a
patient's covariate vector $\mathbf{Z}$, we have \textit{covariate-adaptive
randomization}. In particular, the randomization procedure can then
be described by
$\phi_{j+1}=\Pr(T_{j+1}=1|\mathbf{T}_j,\mathbf{Z}_1,\ldots,\mathbf{Z}_{j+1})$,
noting that the current patient is randomized based on the history of
previous treatment assignments, the covariate vectors of past
patients \textit{and} the current patient's covariate vector. The goal
of covariate-adaptive randomization is to adaptively balance the
covariate profiles of patients randomized to treatments 1 and 2.
Most techniques for doing so have focused on minimizing the
differences of numbers on treatments $1$ and $2$ across strata,
often marginally. Note that covariate-adaptive randomization
induces a complex covariance structure, given by
$\operatorname{Var}(\mathbf{T}_n|\mathbf{Z}_1=\mathbf{z}_1,\ldots,\mathbf{Z}_n
=\mathbf{z}_n)=\bolds{\Sigma}_{n,\mathbf{z}}$.

For a small set of known discrete covariates, prestratification is
the most effective method for forcing balance with respect to those
covariates across the treatment groups. The technique of
prestratification uses a separate restricted randomization
procedure within each stratum. For notational purposes, if discrete
covariate $Z_i, i=1,\ldots,K$, has $k_i$ levels, then restricted
randomization is used within each of the $\prod_{i=1}^K k_i$ strata.

The first covariate-adaptive randomization procedures were proposed
in the mid-1970s. Taves (\citeyear{tav1974}) proposed a deterministic method to
allocate treatments designed to minimize imbalances on important
covariates, called the \textit{minimization} method.\break Pocock and Simon
(\citeyear{pocsim75}) and Wei (\citeyear{wei1978})
described generalizations of minimization to
randomized clinical trials. We will refer to this class of
covariate-adaptive randomization procedures as \textit{marginal}
procedures, as they balance on covariates marginally, within each of
$\sum_{i=1}^K k_i$ levels of given covariates.

The general marginal procedure can be described as follows for a
two-treatment clinical trial. Let\break $N_{ijl}(n)$ be the number of
patients on treatment $l$ in level $j$ of covariate $Z_i$,
$i=1,\ldots,K, j=1,\ldots,k_i, l=1,2$, after $n$ patients have been
randomized. When patient $n+1$ is ready for randomization, the
patient's baseline covariate vector $(Z_1,\ldots,Z_K)$ is observed as
$(z_1,\ldots,z_K)$. Then $D_i(n)=N_{iz_i1}(n) -\break N_{iz_i2}(n)$ is
computed for each $i=1,\ldots,K$. A weighted sum is then taken as
$D(n)=\sum_{i=1}^K w_i D_i(n)$. The measure $D(n)$ is used to
determine the treatment of patient $n+1$. If $D(n)>0$ ($<0$),
then one decreases (increases) the probability of being assigned to
treatment $1$
accordingly. %The method is \textit{marginal} because the difference
%metric is taken as the weighted sum over covariates.
Pocock and Simon (\citeyear{pocsim75})
formulated a general rule using Efron's
(\citeyear{efr1971}) biased coin design as:
\begin{eqnarray*}
\phi_{n+1} & = & \cases{%
1/2, & if $D(n)=0$, \cr
p, & if $D(n)<0$, \cr
1-p, & if $D(n)>0$.}
\end{eqnarray*}
When $p=1$, we have Taves's (\citeyear{tav1974}) minimization method, which is
nonrandomized. Pocock and Simon (\citeyear{pocsim75}) investigated $p=3/4$. %Note
%that this is a covariate-adaptive randomization procedure because
%$\phi_n$ depends on $D(n)$, which in turn depends not only on
%$\boldZ_1,\ldots,\boldZ_n$, but also on $\boldZ_{n+1}$.

Wei (\citeyear{wei1978}) proposed a different marginal procedure using urns. At
the beginning of the trial, each of $\sum_{i=1}^K k_i$ urns contain
$\alpha_1$ balls of type 1 and $\alpha_2$ balls of type 2. Let
$U_{ij}$ denote the urn representing level $j$ of covariate $z_i$,
and let $Y_{ijk}(n)$ be the number of balls of type $k$ in urn
$U_{ij}$ after $n$ patients have been randomized. For each urn
compute the imbalance
$D_{ij}(n)=(Y_{ij1}(n)-Y_{ij2}(n))/(Y_{ij1}(n)+Y_{ij2}(n))$. Suppose
patient $n+1$ has covariate vector $(z_1,\ldots,z_K)$. Select the urn
such that $D_{iz_i}(n)$ is maximized. Draw a ball and replace. If
it is a type $k$ ball, assign the patient to treatment $k$, and add
$\alpha_k$ balls of type $k$ with $\beta_k\ge0$ balls of the
opposite type to each of the observed urns. The procedure is
repeated for each new eligible patient entering the trial. Wei
proved that if there is no interaction between the covariates or
between the treatment effect and covariates in a standard linear
model, then marginal balance is sufficient to achieve an unbiased
estimate of the treatment difference. Efron (\citeyear{efr1980}) provided a
covariate-adaptive randomization procedure that balances both
marginally and within strata, but the method applies only to two
covariates.

There has been substantial controversy in the literature as to
whether the introduction of randomization is necessary when
covariate-adaptive procedures are used. Randomization mitigates the
probability of selection bias and accidental bias, and provides a basis
for inference (e.g., Rosenberger and Lachin, \citeyear{rosLac2002}). Taves's
original paper did not advocate randomization, and, in fact, he
still supports the view that randomization is unnecessary, writing
in a letter to the editor (Taves, \citeyear{tav2004}, page~180):

\begin{quote}
I hope that the day is not too far distant
when we look back on the current belief that randomization is
essential to good clinical trial design and realize that it was\ldots\
``credulous idolatry.''
\end{quote}
Other authors have argued for using minimization without
the additional component of randomization. Aickin (\citeyear{aic2001})
argued
that randomization is not needed in covariate-adaptive procedures
because the covariates themselves are random, leading to randomness
in the treatment assignments. He also argued that the usual
selection bias argument for randomization is irrelevant in
double-masked clinical trials with a central randomization unit.

Several authors, such as Zelen (\citeyear{zel1974}),
Nordle and Brandmark (\citeyear{nordBra1977}),
Efron (\citeyear{efr1980}),
Signorini et al. (\citeyear{sigLeuSimBelGeb93}) and
Heritier, Gebski and Pillai (\citeyear{herGebPil2005}),
proposed covariate-adaptive randomization procedures which
achieve balanced allocation both within margins of the chosen
factors and within strata. These methods emphasize the importance of
balancing over interactions between factors when such exist.\break
Raghavarao (\citeyear{rag1980}) proposed an allocation procedure based on distance
functions. When the new patient enters the trial, one computes
$d_k$, the Mahalanobis distance between the covariate profile of the
patient and the average of the patients already assigned to
treatment~$k$, where $k=1,\ldots,K$. Then the patient is assigned to
treatment~$k$ with probability $p_k\propto d_k$.

%s3 ###
\section{Randomization-Based Inference}\label{s3}

One of the benefits of randomization is that it provides a basis for
inference (see Chapter 7 of Rosenberger and Lachin, \citeyear{rosLac2002}). Despite
this, assessment of treatment effects in clinical trials is often
conducted using standard likelihood-based methods that ignore the
randomization procedure used. Letting $\mathbf{Y}^{(n)}=(Y_1,\ldots,Y_n)$
be the response vector, $\mathbf{T}_n^{(n)}=(T_1,\ldots,T_n)$ the treatment
assignment vector and
$\mathbf{Z}^{(n)}=(\mathbf{Z}_1,\ldots,\mathbf{Z}_n)$ the
covariate vectors of patients $1,\ldots,n$, the likelihood can simply
be written as\looseness=1
\begin{eqnarray*}
\mathcal{L}_n&=&\mathcal{L}
\bigl(\mathbf{Y}^{(n)},\mathbf{T}^{(n)},\mathbf{Z}^{(n)};\theta\bigr)
\\
&=&\mathcal{L}\bigl(Y_n|\mathbf{Y}^{(n-1)},\mathbf{T}^{(n)},
\mathbf{Z}^n;\theta\bigr)
\\
&&{} \cdot \mathcal{L}\bigl(T_n|\mathbf{Y}^{(n-1)},{\mathbf{T}}^{(n-1)},
\mathbf{Z}^{(n)};\theta\bigr)
\\
&&{} \cdot \mathcal{L}\bigl(\mathbf{Z}_n|\mathbf{Y}^{(n-1)},
\mathbf{T}^{(n-1)},\mathbf{Z}^{(n-1)}\bigr)\mathcal{L}_{n-1}.
\end{eqnarray*}
As $\mathcal{L}(Y_n|\mathbf{Y}^{(n-1)},\mathbf{T}^{(n)},
\mathbf{Z}^n;\theta)=\mathcal{L}(Y_n|T_n, \mathbf{Z}_n;\theta)$,
the treatment assignments do not
depend on $\theta$, and the covariates are considered i.i.d.,
we can reduce this to the recursion
\begin{eqnarray*}
\mathcal{L}_n&\propto&\mathcal{L}(Y_n|T_n, \mathbf{Z}_n;\theta)\mathcal{L}_{n-1}
\\
&=& \prod_{i=1}^n \mathcal{L}(Y_i|T_i,\mathbf{Z}_i;\theta).
\end{eqnarray*}
This is the standard regression equation under a population model;
that is, the randomization is ancillary to the likelihood. Thus,
a proponent of the likelihood principle would ignore the design in
the analysis, and proceed with tests standardly available in SAS.

The alternative approach is to use a randomization test, which is a
simple nonparametric alternative. Under the null hypothesis of no
treatment effect, the responses should be a deterministic sequence
unaffected by the treatment assigned. Therefore, the distribution
of the test statistic under the null hypothesis is computed with
reference to all possible sequences of treatment assignments under
the randomization procedure.

Various authors have struggled with the appropriate way to perform
randomization tests following covariate-adaptive randomization.
Pocock and Simon (\citeyear{pocsim75}) initially suggested that the sequence of
covariate values and responses be treated as deterministic, and the
sequence of treatment assignments be permuted for those specific
covariate values. This is the approach taken by most authors.
Ebbutt et al. (\citeyear{ebbKayMcEng97}) presented an example where results
differed when the randomization test took into consideration the
sequencing of patient arrivals. Senn concluded from this that the
disease was changing in some way through the course of the trial and
thus there was a time trend present (see the discussion of Atkinson,
\citeyear{atk1999}).

%s4 ###
\section{What We Know About Covariate- Adaptive Randomization Procedures}\label{s4}

Our knowledge of covariate-adaptive randomization comes from (a) the
original source papers; (b) a vast number of simulation papers; (c)
advocacy or regulatory papers (for or against); and (d) review
papers. Very little theoretical work has been done in this area,
despite the proliferation of papers. The original source papers are
fairly uninformative about theoretical properties of the procedures.
In Pocock and Simon (\citeyear{pocsim75}), for instance, there is a small
discussion, not supported by theory, on the appropriate selection of
biasing probability $p$. There is no discussion about the effect of
the choice of weights for the covariates; no discussion about the
effect on inference; no theoretical justification that the procedure
even works as intended: Do covariate imbalances (loosely defined)
tend to zero? Does marginal balance imply balance within strata or
overall? Wei (\citeyear{wei1978}) devotes less than one page to a description of
his procedure; he does prove that marginal balance implies balance
within strata for a linear model with no interactions. Taves (\citeyear{tav1974})
is a nontechnical paper with only intuitive justification of the
method. Simulation papers have been contradictory.

Klotz (\citeyear{klo1978}) formalized the idea of finding an optimal value of
biasing probability $p$ as a constrained maximization problem.
Consider a trial with $K$ treatments and covariates. When patient
$n+1$ is ready to be randomized, one computes $D_k$, the measure of
overall covariate imbalance if the new patient is assigned to
treatment $k=1,\ldots,K$. The goal is to find the vector of
randomization probabilities $\bolds{\rho}=(\rho_1,\ldots,\rho_K)$ which
maximizes the entropy measure subject to the constraint on the
expected imbalance. Titterington (\citeyear{tit1983}) built upon Klotz's idea and
considered minimization of the quadratic distance between $\bolds{\rho}$
and the vector of uniform probabilities $\bolds{\rho}_0=(1/K,\ldots,1/K)$
subject to the constraints on the expected imbalance.

Aickin (\citeyear{aic2001}) provides perhaps one of the few theoretical analyses
of covariate-adaptive randomization procedures. He gives a very
short proof contradicting some authors' claims that
covariate-adaptive randomization can promote imbalances in
unmeasured covariates. If $X_2$ is an unmeasured covariate, and
covariate-adaptive randomization was used to balance on covariate
$X_1$, then $X_2$ can be decomposed into its \mbox{linear} \mbox{regression} part,
given by $L(X_2|X_1)$, and its linear regression residual
$X_2-L(X_2|X_1)$. If $X_1$ and $X_2$ are correlated positively or
negatively, balancing on $X_1$ will improve the balance of
$L(X_2|X_1)$. Since the residual is not correlated with the
randomization procedure, $X_2-L(X_2|X_1)$ will balance as well as
with restricted or complete randomization. This is a formal
justification of the intuitive argument that Taves (\citeyear{tav1974}) gave in
his original paper, an argument that Aickin (\citeyear{aic2001}) says
is a ``remarkably insightful observation.'' Aickin also uses causal
inference modeling to show that, if the unobserved errors correlated
with the treatment assignments and known covariates are linearly
related to the known covariates, the treatment effect should be
unbiased.

There seems to be a troubling misconception in the literature with
regard to covariate-adaptive randomization. For example, in an
editorial in the \textit{British Medical Journal} (Treasure and MacRae,
\citeyear{treMac1998}) we have the statement:

\begin{quote}
The theoretical validity of the method of
minimisation was shown by Smith$\ldots.$
\end{quote}
The quotation refers to Smith (\citeyear{smi84b}), which actually
derives the asymptotic distribution of the randomization test
following a model-based optimal design approach favored by many
authors. We shall discuss this approach momentarily, but it is
important to point out that \textit{there is no justification},
\textit{theoretical or otherwise},
\textit{of minimization methods in Smith's paper}.

In contrast to the dearth of publications exploring
covariate-adaptive randomization from a theoretical perspective, a
literature search revealed about 30 papers reporting results of
simulation studies. Some of these papers themselves are principally
a review of various other simulation papers. A glance at the recent
Society for Clinical Trials annual meeting abstract guide revealed
about 10 contributed talks reporting additional simulation results
and their use in clinical trials, indicating the continuing
popularity of these designs.

Papers dealing with the comparison of stratified block designs with
covariate-adaptive randomization methods with respect to achieving
balance on covariates include the original paper of Pocock and Simon
(\citeyear{pocsim75}),
Therneau (\citeyear{ther1993}), and review papers by
Kalish and Begg (\citeyear{kalBeg1985}) and
Scott et al. (\citeyear{scoMcpRamCam02}).
The general consensus is that covariate-adaptive
randomization does improve balance for large numbers of covariates.

Inference following covariate-adaptive randomization has been
explored by simulation in Birkett (\citeyear{bir1985}),
using the $t$-test, Kalish and Begg (\citeyear{kalBeg1987})
using randomization tests, and Frane (\citeyear{fra1998}), using
analysis of covariance. Recent papers by
Tu, Shalay and Pater (\citeyear{tuShaPat2000})
and McEntegart (\citeyear{mce2003}) cover a wide-ranging number of questions. Tu
et al. found that minimization method is inferior to
stratification in reducing error rates, and argued that marginal
balance is insufficient in the presence of interactions. McEntegart
concluded that there is little difference in power between
minimization method and stratification. Hammerstrom (\citeyear{ham2003}) performed
some simulations and found that covariate-adaptive randomization
does not significantly improve error rates, but does little harm,
and therefore is useful only for cosmetic purposes.

We conclude this section by interjecting some relevant questions.
Does marginal balance improve power and efficiency, or is it simply
cosmetic? Is covariate-adaptive randomization the proper approach
to this problem?

%s5 ###
\section{Model-Based Optimal Design~Approaches}\label{s5}

An alternate approach to balance is to find the optimal design that
minimizes the variance of the treatment effect in the presence of
covariates. This approach is first found in Harville (\citeyear{har1974}), not in
the context of clinical trials, and in Begg and Iglewicz (\citeyear{begIgl1980}).
The resulting designs are deterministic.

Atkinson (\citeyear{atk1982}) adopted the approach and has advocated it in a
series of papers, and in the 1982 paper, introduced randomization
into the solution. In order to keep consistency with the original
paper, we summarize Atkinson's approach for a general case of $K\ge
2$ treatments. Suppose $K$ treatments are to be compared, and
responses follow the classical linear regression model given by
\begin{eqnarray*}
E(Y_i)=\mathbf{x}_i^{\prime} \bolds{\beta},\quad  i=1,\ldots,n,
\end{eqnarray*}
where the $Y_i$'s are independent with
$\operatorname{Var}({\mathbf{Y}})=\sigma^2{\mathbf{I}}$ and
$\mathbf{x}_i$ is $(K+q)\times1$ vector which includes treatment indicators and
selected covariates of interest ($q$ is the number of covariates in
the model). Let $\hat{\bolds{\beta}}$ be the least squares estimator of
$\bolds{\beta}$. Then
$\operatorname{Var}(\hat{\bolds{\beta}})=\sigma^2({\mathbf{X}}^{\prime}{\mathbf{X}})^{-1}$,
where ${\mathbf{X}}^{\prime}{\mathbf{X}}$ is the dispersion matrix
from $n$ observations.

For the construction of optimal designs we wish to find the $n$
points of experimentation at which some function is optimized (in
our case we will be finding the optimal sequence of $n$ treatment
assignments). The dispersion matrix evaluated at these $n$ points
is given by $\mathbf{M}(\xi_n)={\mathbf{X}}^{\prime}{\mathbf{X}}/n$, where
$\xi_n$ is the $n$-point design. It is convenient, instead of
thinking of $n$ points, to formulate the problem in terms of a
measure $\xi$ (which in this case is a frequency distribution) over a
design region $\Xi=\{1,\ldots,K\}$.

Atkinson formulated the optimal design problem as a design that
minimizes, in some sense, the variance of
${\mathbf{A}}^{\prime}{\hat{\bolds{\beta}}}$, where $\mathbf{A}$ is a matrix
of contrasts. One possible criterion is Sibson's (\citeyear{sib1974})
$D_A$-optimality that maximizes
%
%e1 ###
\begin{eqnarray}\label{aaaa}
|{\mathbf{A}}^{\prime}{\mathbf{M}}^{-1}(\xi){\mathbf{A}}|^{-1}.
\end{eqnarray}

For any multivariable optimization problem, we compute the
directional derivative of the criterion. In the case of the $D_A$
criterion in (\ref{aaaa}), we can derive the Fr\`{e}chet derivative
as
\[
d_A({\mathbf{x}},\xi)={\mathbf{x}}^{\prime}{\mathbf{M}}^{-1}
(\xi){\mathbf{A}}({\mathbf{A}}^{\prime}{\mathbf{M}}^{-1}(\xi)
{\mathbf{A}})^{-1}{\mathbf{A}}^{\prime}{\mathbf{M}}^{-1}(\xi)
{\mathbf{x}},
\]
for $x\in\Xi$. By the classical Equivalence theorem of
Kiefer and Wolfowitz (\citeyear{kieWol1960}), the optimal design $\xi^*$ that
maximizes the criterion (\ref{aaaa}) then satisfies the following
equations:
\begin{eqnarray*}
\sup_{{\mathbf{x}}\in\Xi} d_A({\mathbf{x}},\xi) \le s \quad \forall
\xi\in\Xi
\end{eqnarray*}
and
\begin{eqnarray*}
\sup_{{\mathbf{x}}\in\Xi} d_A({\mathbf{x}},\xi^*)=s.
\end{eqnarray*}

Such a design is optimal for estimating linear contrasts of
$\bolds{\beta}$. Assume $n$ patients have already been allocated, and
the resulting $n$-point design is given by $\xi_n$. Let the value of
$d_A(x,\xi)$ for allocation of treatment $k$ be $d_A(k,\xi)$.
Atkinson proposed a sequential design which allocates the $(n+1)$th
patient to the treatment $k=1,\ldots,K$ for which $d_A(k,\xi_n)$ is a
maximum, given the patient's covariates. The resulting design is
deterministic.

In order to randomize the allocation, Atkinson suggested biasing a
coin with probabilities
%
%e2 ###
\begin{eqnarray}\label{eeee}
\rho_k=\frac{\psi(d_A(k,\xi_n))}{\sum_{k=1}^K\psi(d_A(k,\xi_n))},
\end{eqnarray}
where $\psi(x)$ is any monotone increasing function, and allocating
to treatment $k$ with the corresponding probability. With two
treatments, $k=1,2$, %$(1=A, 2=B)$
we have $s=1$, ${\mathbf{A}}^{\prime}=(-1,1,0,\ldots,0)$, and the
probability of assigning treatment $1$ is given by
%
%e3 ###
\begin{eqnarray}\label{ffff}
\phi_{n+1}=\frac{\psi(d_A(1,\xi_n))}{\psi(d_A(1,\xi_n))
+ \psi(d_A(2,\xi_n))}.
\end{eqnarray}
(We consider only the case of two treatments in this paper.)
Equation (\ref{ffff}) gives a broad class of covariate-adaptive
randomization procedures. The choice of function $\psi$ has not been
explored adequately. Atkinson (\citeyear{atk1982}) suggested using $\psi(x)=x$;
Ball, Smith and Verdinelli (\citeyear{balSmiVer1993})
suggested $\psi(x)=(1+x)^{1/\gamma}$ for a
parameter $\gamma\ge0$, which is a compromise between randomness
and efficiency.

Atkinson (\citeyear{atk1999}, \citeyear{atk2002})
performed careful simulation studies to
compare the performance of several covariate-adaptive randomization
procedures for a linear model with constant variance and trials up
to $n=200$ patients. One criterion of interest was \emph{loss}, the
expected amount of information lost due to treatment and covariate
imbalance. Another criterion was selection bias, measuring
the probability of correctly guessing the next treatment assignment.
Atkinson observed that the deterministic procedure based on the
$D_A$-optimality criterion has the smallest value of loss, and
Atkinson's randomized procedure (\ref{ffff}) with \mbox{$\psi(x)=x$}
increases the loss. He noted that $D_A$-optimal designs are
insensitive to correlation between the covariates, while complete
randomization and minimization method increase the loss when
covariates are correlated.

%s6 ###
\section{What We Know About Atkinson's Class of Procedures}\label{s6}

Considerably more theoretical work has been done on the class of
procedures in (\ref{ffff}) than for the covariate-adaptive
randomization procedures in Section~\ref{s2}. Most of the work has been
done in a classic paper by Smith (\citeyear{smi84a}), although he dealt with a
variant on the procedure in (\ref{ffff}). It is instructive to
convert to his notation:
\[
E(Y_n)=\alpha t_n+\sum_{j=1}^q z_{nj}\beta_j,
\]
where $Y_n$ and $t_n$ are the response and treatment assignments of
the $n$th patient, respectively, and $z_{nj}$ represent~$q$
covariates, and may include an intercept. Let $\mathbf{T}_n$ be the
treatment assignment vector and let $\mathbf{Z}_n$ be the matrix of
covariates. Then Atkinson's procedure in (\ref{ffff}) can be
formulated as follows: assign $t_{n+1}=\pm1$ with probabilities
proportional to $(\pm1-\mathbf{z}_{n+1}'(\mathbf{Z}_n\mathbf{Z}_n)^{-1}
\mathbf{Z}_n\mathbf{t}_n)^2$ (Smith, \citeyear{smi84b}, page~543).
Smith (\citeyear{smi84a}) introduced a more general class of
allocation procedures given by
%
%e4 ###
\begin{equation}\label{2-Smith_allocation_general}
\phi_{n+1}=\psi(n^{-1}\mathbf{z}'_{n+1}\mathbf{Q}^{-1}
\mathbf{Z}_n'\mathbf{t}_n),
\end{equation}
where $\psi$ is nonincreasing, twice continuously differentiable
function with bounded second derivative satisfying
$\psi(x)+\psi(-x)=1$, and
$\mathbf{Q}=E(\mathbf{z}_n\mathbf{z}_n')=\lim_{n\rightarrow
\infty}n^{-1}(\mathbf{Z}_n'\mathbf{Z}_n)$.
It is presumed that the $\{\mathbf{z}_n\}$ are independent, identically
distributed random vectors, $\mathbf{Q}$~is nonsingular and all third
moments of $\mathbf{z}_n$ are finite. Note that the procedure
(\ref{2-Smith_allocation_general}) can be implemented only if the
distribution of covariates is known in the beginning of
the trial. %Smith then explores a variant of (\ref{ffff}), given by
%$$\phi_n=\psi (\boldz_{n+1} (\underset{n\to\infty}{\lim}(
% ),$$
%where $\psi$ is non-increasing and $\psi(x)=\psi(-x)=1$. It is
%presumed that the limiting term converges to the underlying
%population mean of the covariates, and thus the procedure can only
%be implemented in practice at that value.

Smith suggested various forms of $\psi$, most leading to a
proportional biased coin raised to some power $\rho$. In general,
$\rho=-2\psi^{\prime}(0)$. Without covariates, Atkinson's procedure
in (\ref{eeee}) leads to
\[
\phi_{n+1}= \frac{n_2^{\rho}}{n_1^{\rho}+n_2^{\rho}},
\]
where
$\rho=2$. Smith found the asymptotic variance of the randomization
test based on the simple treatment effect, conditional on
$\mathbf{Z}_n$. He did not do any further analysis or draw conclusions
except to suggest that $\rho$ should be selected by the investigator
to be as large as possible to balance the competing goals of
balance, accidental bias and selection bias.

%s7 ###
\section{Balance, Efficiency or Ethics?}\label{s7}

%--------------------
%f1 ###
\begin{figure}[b]

\includegraphics{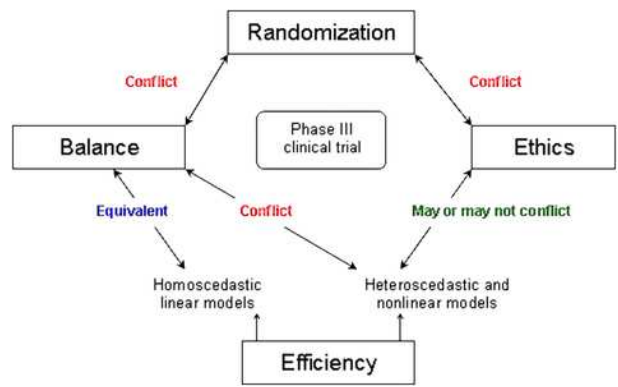}

\caption{Multiple objectives of a phase \textup{III} clinical
trial.\label{Multiple objectives}}
\end{figure}
%%--------------------

Clinical trials have multiple objectives. The principal
considerations are given in the schematic in Figure \ref{Multiple
objectives}. \textit{Balance} across treatment groups is often
considered essential both for important covariates and for treatment
numbers themselves. \textit{Efficiency} is critical for demonstrating
efficacy. \textit{Randomization} mitigates certain biases.
\textit{Ethics} is an essential component in any human experimentation, and
dictates our treatment of patients in the trial. These
considerations are sometimes compatible, and sometimes in conflict.
In this section, we describe the interplay among balance,
efficiency and ethics in the context of randomized clinical trials,
and give some examples where they are in conflict.

In a normal error linear model with constant variance, numerical
balance between treatments on the margins of the covariates is
equivalent to minimizing the variance of the treatment effect. This
is not true for nonlinear models, such as logistic regression or
traditional models for survival analysis (Begg and Kalish, \citeyear{begKal1984};
Kalish and Harrington, \citeyear{kalHar1988}). As we shall discuss further in the
next section, balance does not imply efficiency except in
specialized cases. This leaves open the question, is balance on
covariates important?

We have the conflict recorded in a fascinating interchange among
Atkinson, Stephen Senn and John Whitehead (Atkinson, \citeyear{atk1999}).
Whitehead argues:

\begin{quote}
I think that one criterion is really to
reduce the probability of some large imbalance rather than the
variance of the esti\-mates$\ldots.$ And to make sure that these
unconvincing trials, because of the large imbalance, happen with
very low probability, perhaps is more important$\ldots.$ I would always
be wanting to adjust for these variables. None the less, the message
is simpler if my preferred adjusted analysis is similar to the
simple message of the clinicians.
\end{quote}
Senn gives the counterargument:

\begin{quote}
I think we should avoid pandering to these
foibles of physicians$\ldots.$ I think people worry far too much about
imbalance from the inferrential (sic) point of view$\ldots.$ The way I
usually describe it to physicians is as follows: if we have an
unbalanced trial, you can usually show them that by throwing away
some patients you can reduce it to a perfectly balanced trial. So
you can actually show that within it there is a perfectly balanced
trial. You can then say to them: `now, are you prepared to make an
inference on this balanced subset within the trial?' and they nearly
always say `yes.' And then I~say to them, `well how can a little
bit more information be worse than having just this balance trial
within it?'
\end{quote}

We thus encounter once again deep philosophical differences and the
ingrained culture of clinical trialists. Fortunately, balance and
efficiency are equivalent in homoscedastic linear models. Thus,
stratified randomization and covariate-adaptive randomization
procedures (such as Pocock and Simon's\break method) are valid to the
degree in which they force balance over covariates. Atkinson's
model-based approach is an alternative method that can incorporate
treatment-by-covariate interactions and continuous covariates.
Atkinson's class of procedures for linear models has an advantage of
being based on formal optimality criteria as opposed to ad hoc
measures of imbalance used in covariate-adaptive randomization
procedures. On the other hand, balanced designs may not be most
efficient in the case of nonlinear and heteroscedastic models. We
agree with Senn that cosmetic balance, while psychologically
reassuring, should not be the goal if power or efficiency is lost in
the process of forcing balance.

%In clinical trials with binary or survival time outcomes the most
%efficient design will depend on unknown model parameters, which must
%be sequentially estimated. The introduction of nonlinearity into the
%problem also leads to the development of allocation methods which
%are appealing for ethical reasons. Note that the most efficient
%designs, such as Atkinson's procedures, can result in allocations
%where more patients are assigned to the inferior treatment.

First, let us illustrate that balanced allocation can be less
efficient and less ethically appealing than unbalanced allocation in
some instances, and that there may exist unbalanced designs which
outperform balanced designs in terms of compound objectives of
efficiency and ethics. Consider a binary response trial of size $n$
comparing two treatments $A$ and $B$, and suppose there is an
important binary covariate $Z$, say gender ($Z=0$ if a patient is
male, and $Z=1$ if female), such that there are $n_0$ males and
$n_1$ females in the trial. Also assume that success probabilities
for treatment $k$ are $p_{k0}$ for males and $p_{k1}$ for females,
where $k=A,B$. Let $q_{kj}=1-p_{kj}$, $j=0,1$. For the time being we
will assume that the true success probabilities are known. One
measure of the treatment effect for binary responses is the log-odds
ratio, which can be expressed as
%
%e5 ###
\begin{equation}\label{2_log-OR}
\log\operatorname{OR}(Z=j)=\log\frac{p_{Aj}/q_{Aj}}{p_{Bj}/q_{Bj}}, \quad  j=0,1.
\end{equation}

An experimental design question is to determine allocation proportions
$\pi_{Aj}$ and $\pi_{Bj}$ in stratum $j$ for treatments $A$ and $B$,
respectively, where $j=0$ (male) or $j=1$ (female). Let us consider the
following three allocation rules:
\begin{longlist}[Rule 3:]
\item[\textit{Rule} 1:] Balanced treatment assignments in the two strata,
given by
\[
\pi_{Aj}=\pi_{Bj}=1/2, \quad j=0,1;
\]
\item[\textit{Rule} 2:] Neyman allocation maximizing the power of the stratified
asymptotic test of the log-odds ratio:
\[
T_j=\frac{\log\widehat{\operatorname{OR}}(Z=j)}
{\sqrt{\widehat{\operatorname{var}} (\log\widehat{\operatorname{OR}}(Z=j) )}},
\quad j=0,1.
\]
The allocation proportion is given by
\[
\pi^*_{Aj}=\frac{1/\sqrt{p_{Aj}q_{Aj}}}{1/\sqrt
{p_{Aj}q_{Aj}}+1/\sqrt{p_{Bj}q_{Bj}}},\quad  j=0,1;
\]
\item[\textit{Rule} 3:] the analog of
Rosenberger et al.'s (\citeyear{rosStaIvaHarRic01}) optimal
allocation minimizing the expected number of treatment failures in the
trial subject to the fixed variance of the log-odds ratio. This is
given by
\[
\pi^{**}_{Aj}=\frac{1/\sqrt{p_{Aj}q_{Aj}^2}}
{1/\sqrt{p_{Aj}q_{Aj}^2}+1/\sqrt{p_{Bj}q_{Bj}^2}}, \quad j=0,1.
\]
\end{longlist}

Note that unlike Rule 1, Rules 2 and 3 depend on success probabilities
in the two strata, and are unbalanced, in general. Consider a case when
$n_0=n_1=100$ and let $(p_{A0},p_{B0})=(0.95,0.7)$ and
$(p_{A1},\break p_{B1})=(0.7,0.95)$. This represents a case when one of the
treatments is highly successful, there is significant treatment
difference between $A$ and $B$, and there is treatment-by-covariate
interaction (treatment $A$ is more successful for males and is less
successful for females). Then allocation proportions for treatment $A$
in the two strata are $\pi_{A0}=0.68$ and $\pi_{A1}=0.32$ for Rule 2,
and $\pi_{A0}=0.84$ and $\pi_{A1}=0.16$ for Rule 3.

All three rules are very similar in terms of efficiency, as measured
by the asymptotic variances of stratum-specific estimates of the
log-odds ratio. However, Rules 2 and 3 provide extra ethical
savings. For the sample size considered, Rule 3 is expected to have
16 fewer failures than the balanced design. At the same time, Rule
2, whose primary purpose is optimizing efficiency, is expected to
have 8 fewer failures than the balanced allocation. Therefore, in
addition to maximizing efficiency, Rule 2 provides additional
ethical savings, and is certainly far more attractive than balanced
allocation.

So far we have compared different target allocations for ``fixed'' designs,
that is, for a given number of patients in each treatment group and known
model parameters.
In practice, true success probabilities are not available at the trial onset,
which precludes direct implementation of Rules 2 and 3. Since clinical
trials are
sequential in nature, one can use accruing responses to estimate the parameters,
and then cast a randomization procedure which asymptotically achieves
the desired allocation.
%Such an approach induces additional variability of allocation
%probabilities, which may reduce power.
To study operating characteristics of response-adaptive
randomization procedures targeting Neyman allocation (Rule 2) and
optimal allocation (Rule 3) we ran a simulation study in R using
10,000~replications (results are available from the second author upon
request). In the
simulations we assumed that two strata (male and female) are equally
likely. For Rules 2 and 3, the doubly adaptive biased coin design
(DBCD) procedure of Hu and Zhang (\citeyear{huZha2004}) was used within each stratum
to sequentially allocate patients to treatment groups. In addition,
balanced allocation was implemented using stratified permuted block
design (PBD) with block size $m=8$. We assumed that responses are
immediate, and compared the procedures with respect to power of the
stratified asymptotic test of the log-odds ratio for testing the
null hypothesis $H_0$: $(p_{A0}=p_{B0})$ and $(p_{A1}=p_{B1})$ versus
$H_A$: not $H_0$ using significance level $\alpha=0.05$, and the
expected number of treatment failures. We considered several
experimental scenarios for success probabilities
$(p_{Aj},p_{Bj})$, $j=0,1$, including the one described in the example above.
To facilitate comparisons, the sample size for each experimental
scenario was chosen such that the stratified block design achieves
approximately 80$\%$ power of the test. In summary,
response-adaptive randomization procedures worked as expected: for
chosen sample sizes they converged to the targeted allocations and
preserved the nominal significance level. Additionally,
response-adaptive randomization procedures had similar average power
to the PBD, but on average they had fewer treatment failures.
Ethical savings of response-adaptive designs were more pronounced
when one of the treatments had high success probability (0.8--0.9)
and treatment differences were large.

We would also like to emphasize that phase III trials are pivotal
studies, and one typically has an idea about the success probabilities
of the treatments from early stage trials. If a particular allocation
is such that it leads to high power of the test, and it is also skewed
toward the better treatment, then it makes sense to implement such a
procedure. The additional ethical savings can be prominent if the
ethical costs associated with trial outcomes are high, such as deaths
of trial participants.

%s8 ###
\section{CARA Randomization}\label{s8}

Hu and Rosenberger (\citeyear{huRos2006}) define a covariate-adjusted
response-adaptive (CARA) randomization procedure as one for which
randomization probabilities for a current patient depend on the
history of previous patients' treatment assignments, responses and
covariates, and the covariate vector of the current patient, that is,
%
%e6 ###
\begin{equation}\label{2_CARA}
\quad
\phi_{j}=\Pr(T_{j+1}=1|\mathbf{T}_j,\mathbf{Y}_j,
\mathbf{Z}_1,\ldots,\mathbf{Z}_j,\mathbf{Z}_{j+1}).
\end{equation}
There have been only few papers dealing with CARA
randomization, and it has become an area of active research. CARA
randomization is
an extension of \textit{response-adaptive randomization} which deals with
adjustment for covariates.
Response-adaptive randomization has a rich history in the literature,
and the
interested reader is referred to
Section~1.2 of Hu and Rosenberger (\citeyear{huRos2006}).

Bandyopadhyay and Biswas (\citeyear{banBis2001})
considered a linear regression model
for two treatments and covariates with an additive treatment effect and
constant variance. Suppose large values of response correspond to a
higher efficacy. Then the new patient is randomized to treatment $1$
with probability
%
%e7 ###
\begin{eqnarray}\label{BB}
\phi_{j+1}=\Phi(d_j/T),
\end{eqnarray}
where $d_j$ is the difference of covariate-adjusted treatment means
estimated from the first $j$ patients, $T$ is a scaling constant
and $\Phi$ is the standard normal c.d.f. Although procedure
(\ref{BB}) depends on the full history from $j$ patients, it does
not account for covariates of the $(j+1)$th patient, and it is not
a CARA procedure in the sense of (\ref{2_CARA}). Also, this
procedure depends on the choice of $T$, and small values of $T$ can
lead to severe treatment imbalances which can lead to high power
losses.

Atkinson and Biswas (\citeyear{atkBis05a}, \citeyear{atkBis05b})
improved the allocation rule of
Bandyopadhyay and Biswas (\citeyear{banBis2001})
by proposing CARA procedures that are
based on a weighted $D_A$-optimal criterion combining both efficiency
and ethical considerations. They investigated operating characteristics
of the proposed designs\break through simulation, but they did not derive
asymptotic properties of the estimators and allocation proportions.
Without the asymptotic properties of the estimators, it is difficult to
assess the validity of statistical inferences following CARA designs.

%Biswas, Huang and Huang (2006) proposed another design for continuous
%responses with covariates which is a modification of the
%drop-the-loser urn model by Ivanova (2003). For their design, an urn
%contains balls of three types $A$, $B$, and 1, where 1 is an
%immigration ball. Suppose $j$ patients have been randomized and
%treated, and patient $(j+1)$. enters the trial with covariate $
%is of either type $A$ or $B$, then the patient is assigned to the
%corresponding treatment. If the ball is a type 1 ball, then it is
%replaced into the urn together with one ball of each type $A$ and $B$
%and a drawing procedure is repeated until one gets a type $A$ or $B$
%ball.Let $Y_{j+1}$ be the response of patient $(j+1)$. Then the ball
%is replaced to the urn with probability
%for some constants $c$ and $T$ selected by an experimenter. The
%authors compared design (\ref{BHH}) with (\ref{BB}) through
%simulations and concluded that the latter is superior in terms of
%power and ethics (larger proportion of patients on a better treatment)

A few papers describe CARA designs for binary response trials. One of
the first papers in this field is by Rosenberger, Vidyashankar and
Agarwal (\citeyear{rosVidAga2001}).
They assumed that responses in treatment group $k=A,B$
follow the logistic regression model
\begin{eqnarray*}
\operatorname{logit}\bigl(\Pr(Y_{k}=1|\mathbf{Z}=\mathbf{z})\bigr)
=\bolds{\theta}_k'\mathbf{z},
\end{eqnarray*}
where $\bolds{\theta}_k$ is a vector of model parameters for treatment~$k$. Let
$\hat{\bolds{\theta}}_{jA}$ and $\hat{\bolds{\theta}}_{jB}$ be the maximum
likelihood estimators of model parameters computed from the data from
$j$ patients. Then the $(j+1)$th patient is randomized to treatment
$A$ with probability
\begin{eqnarray*}
\phi_{j+1}=F\bigl((\hat{\bolds{\theta}}_{jA}
-\hat{\bolds{\theta}}_{jB})'\mathbf{z}_{j+1}\bigr),
\end{eqnarray*}
where $F$ is the standard logistic c.d.f. Basically, each patient is
allocated according to the current value of covariate-adjusted odds
ratio comparing treatments $A$ and $B$. The authors compared their
procedure with complete randomization through simulations assuming
delayed responses. They showed that for larger treatment effects
both procedures have similar power, but at the same time the former
results in a smaller expected proportion of treatment failures.

Bandyopadhyay, Biswas and Bhattacharya (\citeyear{banBisBha2007}) also dealt with binary
responses. They proposed a two-stage design for the logistic
regression model. At the first stage, $2m$ patients are randomized
to treatment $A$ or $B$ in a $1\dvtx1$ ratio and accumulated data are
used to estimate model parameters. At the second stage, each patient
is randomized to treatment $A$ with a probability which depends on
the treatment effect estimated from the first stage and the current
patient's covariate vector.

Theoretical properties of CARA procedures have been developed in a
recent paper by Zhang et al. (\citeyear{zhaHuCheChan2007}).
This paper proposed a general framework for CARA randomization
procedures for a very broad class of models,
including generalized linear models. In the paper the authors proved
strong consistency and asymptotic normality
of both maximum likelihood estimators and allocation proportions. They
also examined the CARA design of
Rosenberger, Vidyashankar and Agarwal (\citeyear{rosVidAga2001})
and provided\break asymptotic
properties of the procedure.

CARA procedures do not lend themselves to analysis via
randomization-based inference. The theoretical validity of
randomization tests is based on conditioning on the outcome data as
a set of sufficient statistics, and then permuting the treatment
assignments. Under the null hypothesis of no treatment difference,
the observed outcome data should be exchangeable, leading to a valid
randomization $p$-value (see Pesarin, \citeyear{pes2001}). However, under the
CARA procedure, the treatment assignments and outcomes form the
sufficient statistics, and conditioning on both would leave nothing.
One could perform a standard permutation test on the resulting data
by introducing a ``sham'' equiprobable randomization, but one would
lose information about treatment efficacy.

Therefore, we rely on likelihood-based methods to conduct inference
following a CARA randomization procedure, and Zhang et al.
(\citeyear{zhaHuCheChan2007})
provide the necessary asymptotic theory. For further
discussion of appropriate inference procedure following general
response-adaptive randomization procedures, refer to Chapter 3 of Hu
and Rosenberger (\citeyear{huRos2006}) and
Baldi Antognini and Giovagnoli (\citeyear{baldiGio2005}, \citeyear{baldiGio2006}).

%s9 ###
\section{Comparing Different Randomization Procedures Which Account
for~Covariates}\label{s9}

In the following we used simulation to compare the operating
characteristics of several covariate-adaptive randomization procedures
and CARA procedures for the logistic regression model. We used the
covariate structure considered in Rosenberger, Vidyashankar and Agarwal
(\citeyear{rosVidAga2001}). Assume that responses
for treatment~$k$ satisfy the following logistic regression model:
%
%e8 ###
\begin{equation}\label{7-Logistic_model_Zhang}
\operatorname{logit}\bigl(\Pr(Y_k=1|\mathbf{z})\bigr)
= \alpha_k+\sum_{j=1}^3\beta_{kj}z_j,
\end{equation}
where $\alpha_k$ is the treatment effect, and $\beta_{kj}$ is the
effect due to the $j$th covariate in treatment group $k=A,B$. The
parameter of interest is the covariate-adjusted treatment difference
$\alpha_A-\alpha_B$. The\break components of covariate vector
$\mathbf{z}'=(z_1,z_2,z_3)$,\break which represent \emph{gender}, \emph{age} and
\emph{cholesterol level}, were assumed to be independently distributed
as Bernoulli$(1/2)$, Discrete Uniform$[30,75]$ and\break Normal$(200,20)$. Note
that  model~(\ref{7-Logistic_model_Zhang}) allows for\break
treatment-by-covariate interactions, since covariate
effects $\beta_{kj}$'s are not the same across the treatments.

The operating characteristics of designs included measures of
balance, efficiency and ethics. For \emph{balance} we considered the
allocation proportion\break $N_A(n)/n$, and the allocation proportions
within the male category of covariate \emph{gender},
$N_{A0}(n)/N_0(n)$. Also, we examined the Kolmogorov--Smirnov
distance $d_{\mathrm{KS}}(z_2)$ between empirical distributions of covariate
\emph{age} in treatment groups $A$ and $B$. The \emph{efficiency} of
procedures was measured by the average power of the asymptotic test
of the log-odds ratio evaluated at a given $\mathbf{z}_0$. The
\emph{ethical} aspect of a procedure was assessed by the total
number of treatment failures, $F(n)$.

The sample size $n$ was chosen in such a way that complete
randomization yields approximately 80$\%$ or 90$\%$ power of the test
of log-odds ratio under a particular alternative. For each choice of
$n$ we also estimated the significance level of the test under the null
hypotheses. We report the results for three sets of parameter values
given in Table~\ref{Table7-0}. Under the null hypothesis of no
treatment difference (Model 1), $n=200$. When $\alpha_A-\alpha_B=-1$
(Model 2), the choice of $n=200$ yields $80\%$ power for complete
randomization. When $\alpha_A-\alpha_B=-1.25$ (Model 3), we let
$n=160$, which corresponds to $90\%$ power for complete randomization.

%------------------
%t1 ###
\begin{table}[b]
\tabcolsep=0pt
\caption{Parameter values for the logistic
regression model (\protect\ref{7-Logistic_model_Zhang}) used in~simulations\label{Table7-0}}
\begin{tabular*}{\columnwidth}{@{\extracolsep{\fill}}l d{2.3}d{2.3}d{2.3}d{2.3}d{2.3}d{2.3}@{}}
\hline
& \multicolumn{6}{c@{}}{\textbf{Model}}\\
\cline{2-7}
& \multicolumn{2}{c}{\textbf{1}}
& \multicolumn{2}{c}{\textbf{2}}
& \multicolumn{2}{c@{}}{\textbf{3}}\\
\ccline{2-3,4-5,6-7}
\textbf{Parameters}
& \multicolumn{1}{c}{$\bolds{A}$}
& \multicolumn{1}{c}{$\bolds{B}$}
& \multicolumn{1}{c}{$\bolds{A}$}
& \multicolumn{1}{c}{$\bolds{B}$}
& \multicolumn{1}{c}{$\bolds{A}$}
& \multicolumn{1}{c@{}}{$\bolds{B}$}\\
\hline
$\alpha_k$ & -1.652 & -1.652 & -1.402 & -0.402 & -1.652 & -0.402\\
$\beta_{k1}$ & -0.810 & -0.810 & -0.810 & 0.173 & -0.810 & 0.173 \\
$\beta_{k2}$ & 0.038 & 0.038 & 0.038 & 0.015 & 0.038 & 0.015 \\
$\beta_{k3}$ & 0.001 & 0.001 & 0.001 & 0.004 & 0.001 & 0.004 \\
\hline
\end{tabular*}
\end{table}
%------------------

The first class of procedures are CARA designs. For their
implementation, we need to sequentially estimate model parameters. In
our simulations we assumed that all responses are immediate after
randomization, although we can add a queuing structure to explore the
effects of delayed response. For CARA procedures, some data must
accumulate so that the logistic model is estimable. We used Pocock and
Simon's method to allocate the first $2m_0$ patients to treatments $A$
and $B$.

Suppose after $n>2m_0$ allocations the m.l.e. of $\bolds{\theta}_k$ has
been computed as $\hat{\bolds{\theta}}_{n,k}$. Then, for a sequential
m.l.e. CARA procedure, the $(n+1)$th patient with covariate
$\mathbf{z}_{n+1}$ is allocated to treatment $A$ with probability
$\phi_{n+1}=\rho(\hat{\bolds{\theta}}_{n,A},\hat{\bolds{\theta}}_{n,B},\mathbf{z}_{n+1})$.
We explored four different choices of $\rho$:
\begin{enumerate}
\item Rosenberger, Vidyashankar and Agarwal's (\citeyear{rosVidAga2001})
target:
%
%e9 ###
\begin{eqnarray*}\label{RVA}
\rho_1=\frac{p_A(\mathbf{z})/q_A(\mathbf{z})}{p_A(\mathbf{z})/
q_A(\mathbf{z})+p_B(\mathbf{z})/q_B(\mathbf{z})}.
\end{eqnarray*}
\item Covariate-adjusted version of Rosenberger et al.'s
(\citeyear{rosStaIvaHarRic01}) allocation:
%
%e10 ###
\begin{eqnarray*}\label{RSIHR}
\rho_2&=& \frac{\sqrt{p_A(\mathbf{z})}}{\sqrt{p_A(\mathbf{z})}+\sqrt{p_B(\mathbf{z})}}.
\end{eqnarray*}
\item Covariate-adjusted version of Neyman allocation:
%
%e11 ###
\begin{eqnarray*}\label{Neyman}
\quad
\rho_3=\frac{\sqrt{p_B(\mathbf{z})q_B(\mathbf{z})}}
{\sqrt{p_B(\mathbf{z})q_B(\mathbf{z})}
+\sqrt{p_A(\mathbf{z})q_A(\mathbf{z})}}.
\end{eqnarray*}
\item Covariate-adjusted version of optimal allocation:
%
%e12 ###
\begin{eqnarray*}\label{5-CARA_optimal}
\quad
\rho_4=\frac{\sqrt{p_B(\mathbf{z})}q_B(\mathbf{z})}
{\sqrt{p_B(\mathbf{z})}q_B(\mathbf{z})
+ \sqrt{p_A(\mathbf{z})}q_A(\mathbf{z})}.
\end{eqnarray*}
\end{enumerate}
Here $p_k(\mathbf{z})=1/(1+\exp(-\bolds{\theta}'_k\mathbf{z}))$
and $q_k(\mathbf{z})=1-p_k(\mathbf{z})$, $k=A,B$. We will
refer to CARA procedures with four described targets as \textit{CARA} 1,
\textit{CARA} 2, \textit{CARA}~3 and \emph{CARA} 4, respectively.

We also considered an analogue of Akinson and Biswas's (\citeyear{atkBis05a})
procedure for the binary response case. It is worthwhile to describe
this approach in more detail. Consider model
(\ref{7-Logistic_model_Zhang}) and let $\bolds{\theta}_k=(\alpha_k,\break \beta
_{1k},\beta_{2k}, \beta_{3k})'$. Suppose that a trial has $n_A$
patients allocated to treatment $A$ and $n_B=n-n_A$ patients allocated
to treatment $B$. Then the information matrix about
$\bolds{\theta}=(\bolds{\theta}_A,\bolds{\theta}_B)$ based on $n$
observations is of the form
\[
\mathbf{M}_n={\rm diag}\{\mathbf{Z}_A'\mathbf{W}_A
\mathbf{Z}_A,\mathbf{Z}_B'\mathbf{W}_B\mathbf{Z}_B\},
\]
where $\mathbf{Z}_k$ is the $n_k\times p$ matrix of covariates for
treatment $k$, $\mathbf{W}_k$ is $n_k\times n_k$ diagonal matrix with
elements $p_kq_k$. Here $p_k=p_k(\mathbf{z}_i,\theta_k)$ denote the success
probability on treatment $k$ given $\mathbf{z}_i$ and $q_k=1-p_k$, $k=A,B$.
Suppose the $(n+1)$th patient enters the trial. Then the directional
derivative of the criterion $\det(\mathbf{M})$ for treatment $k$ given
$\mathbf{z}_{n+1}$ is computed as
%
%e13 ###
\begin{equation}\label{Directional derivative}
\qquad
d(k,\bolds{\theta}_n,\mathbf{z}_{n+1})
= \mathbf{z}'_{n+1}(\mathbf{Z}'_k\mathbf{W}_k\mathbf{Z}_k)^{-1}
\mathbf{z}_{n+1}p_kq_k.
\end{equation}

Note that (\ref{Directional derivative}) depends on $\bolds{\theta}_k$,
which must be estimated using the m.l.e. $\hat{\bolds{\theta}}_{n,k}$.
The $(n+1)$th
patient is randomized to treatment $A$ with probability
%
%e14 ###
\begin{equation}\label{7-Atkinson-Biswas}
\phi_{n+1}=\frac{\hat{f}_A d(A,\hat{\bolds{\theta}}_{n,A},
\mathbf{z}_{n+1})}
{\sum_{k=A}^B\hat{f}_k d(k,\hat{\bolds{\theta}}_{n,k},
\mathbf{z}_{n+1})},
\end{equation}
where $f_k$ is the desired proportion on treatment
$k$. We take $f_k=p_k(\mathbf{z})/q_k(\mathbf{z})$. The CARA
procedure (\ref{7-Atkinson-Biswas}) will be referred to as \emph{CARA 5}.

The second class of allocation rules are covariate-adaptive
randomization procedures. For Pocock and Simon's (P--S)
procedure, each component of $\mathbf{z}_{n+1}$ is discretized
into two levels, and the sum of marginal imbalances within these levels
is computed. The $(n+1)$th patient is allocated with probability $3/4$
to the treatment which would minimize total covariate imbalance. If
imbalances for treatments $A$ and $B$ are equal, then the patient is
assigned to either treatment with probability $1/2$.

For the stratified permuted block design (SPBD), the stratum of
the current patient is determined based on the observed combination of
the patient's covariate profile. Within that stratum allocations are
made using permuted blocks of size $m=10$. It is possible that had some
unfilled last blocks, and thus perfect balance is not achieved.
However, we did not specifically examine this feature of SPBD. We also
report the results for complete randomization (CRD).

The program performing the simulations was written in R. For each
procedure, a trial with $n$ patients was simulated 5000 times. To
facilitate the comparison of the procedures, the $n\times4$ matrix of
covariates $\mathbf{Z}$ was generated once and was held fixed for all
simulations. For CARA procedures, the first $2m_0=80$ patients were
randomized by Pocock and Simon's procedure with biasing probability
$p=3/4$. The response probabilities of patients in treatment group
$k=A,B$ were computed by multiplying the rows of $\mathbf{Z}$ by the
vector of model parameters and calculating the logistic c.d.f.
$F(x)=1/(1+\exp(-x))$ at the computed values. The significance level
of the test was set $\alpha=0.05$, two-sided.

%------------------
%t2 ###
\begin{table*}
\caption{Simulation results for
Model 1 with $\bolds{\theta}_A=\bolds{\theta}_B$ and
$n=200$\label{Table7-1}}
\begin{tabular*}{\textwidth}{@{\extracolsep{\fill}}l ccccc@{}}
\hline
\textbf{Procedure} & $\bolds{\frac{N_A(n)}{n}}$ \textbf{(S.D.)}
& $\bolds{\frac{N_{A0}(n)}{N_0(n)}}$ \textbf{(S.D.)}
& $\bolds{d_{\mathrm{KS}}(z_2)}$ \textbf{(S.D.)}
& \textbf{Err. rate}
& $\bolds{F(n)}$ \textbf{(S.D.)} \\
\hline
CRD & 0.50 (0.03) & 0.50 (0.05) & 0.12 (0.04) & 0.05 & 90 (6) \\
SPBD & 0.50 (0.03) & 0.50 (0.04) & 0.12 (0.03) & 0.05 & 90 (6) \\
P--S & 0.50 (0.00) & 0.50 (0.01) & 0.10 (0.03) & 0.05 & 90 (6) \\
CARA 1 & 0.50 (0.03) & 0.50 (0.04) & 0.11 (0.03) & 0.06 & 90 (6) \\
CARA 2 & 0.50 (0.03) & 0.50 (0.04) & 0.12 (0.03) & 0.05 & 90 (6)\\
CARA 3 & 0.50 (0.02) & 0.50 (0.04) & 0.11 (0.03) & 0.06 & 90 (6) \\
CARA 4 & 0.50 (0.02) & 0.50 (0.04) & 0.12 (0.03) & 0.06 & 90 (6) \\
CARA 5 & 0.50 (0.02) & 0.50 (0.04) & 0.12 (0.04) & 0.05 & 90 (6) \\
\hline
\end{tabular*}
\end{table*}
%------------------

Table \ref{Table7-1} shows the results under the null hypothesis
(Model 1). We see that all rules produce balanced allocations. CARA 1,
CARA 3 and CARA 4 procedures are slightly anticonservative, with a
type I error rate of $0.06$, while the procedures CARA 2 and CARA 5
preserve the nominal significance level of $0.05$. Pocock and Simon's
procedure is the least variable among the eight rules considered; the
other procedures are almost identical in terms of variability of
allocation proportions.

%------------------
%t3 ###
\begin{table*}[b]
\caption{Simulation results for
Model 2 with $\alpha_A-\alpha_B=-1.0$ and $n=200$\label{Table7-2}}
\begin{tabular*}{\textwidth}{@{\extracolsep{\fill}}l ccccc@{}}
\hline
\textbf{Procedure}
& $\bolds{\frac{N_A(n)}{n}}$ \textbf{(S.D.)}
& $\bolds{\frac{N_{A0}(n)}{N_0(n)}}$ \textbf{(S.D.)}
& $\bolds{d_{\mathrm{KS}}(z_2)}$ \textbf{(S.D.)}
& \textbf{Power}
& $\bolds{F(n)}$ \textbf{(S.D.)} \\
\hline
CRD & 0.50 (0.04) & 0.49 (0.05) & 0.12 (0.04) & 0.80 & 62 (6) \\
SPBD & 0.50 (0.03) & 0.50 (0.04) & 0.12 (0.03) & 0.81 & 62 (6) \\
P--S & 0.50 (0.01) & 0.50 (0.01) & 0.10 (0.03) & 0.81 & 62 (6) \\
CARA 1 & 0.40 (0.04) & 0.45 (0.04) & 0.12 (0.03) & 0.76 & 56 (6) \\
CARA 2 & 0.48 (0.03) & 0.49 (0.04) & 0.12 (0.03) & 0.81 & 60 (6)\\
CARA 3 & 0.48 (0.03) & 0.49 (0.04) & 0.12 (0.03) & 0.81 & 60 (6) \\
CARA 4 & 0.45 (0.03) & 0.48 (0.04) & 0.12 (0.03) & 0.80 & 58 (6) \\
CARA 5 & 0.47 (0.03) & 0.50 (0.04) & 0.12 (0.04) & 0.81 & 60 (6) \\
\hline
\end{tabular*}
\end{table*}
%------------------

Tables \ref{Table7-2} and \ref{Table7-3} show the results for Models
2~and~3, respectively. The conclusions are similar in the two cases,
and so we will focus on Model 2. Balanced designs equalize the
treatment assignments very well. As expected, the stratified blocks
and Pocock and Simon's procedure are less variable than complete
randomization. Similar conclusions about balancing properties of the
designs apply to balancing with respect to the continuous covariates.
The average power is $90\%$ for the stratified blocks
and Pocock and Simon's procedure, and $89\%$ for complete randomization.

Let us now examine the performance of CARA procedures. All CARA
procedures are more variable than the stratified blocks and Pocock
and Simon's method, but a little less variable than complete
randomization. In addition, all CARA procedures do a good job in
terms of balancing the distributions of the continuous covariates
[estimated $d_{\mathrm{KS}}(z_2)=0.13$ (S.D.${}=0.04$) versus $0.14$ (S.D.${}=0.04$) for
complete randomization]. CARA 2, CARA 3 and CARA~5 procedures are
closest to the balanced design. The simulated allocation proportions
for treatment $A$ and the corresponding standard deviations are $0.48$
(0.03) for CARA 2, and $0.48$ (0.03) for CARA 3, and $0.47$
(0.03) for CARA~5 procedure. These three CARA procedures have average
power of $81\%$, same as for stratified blocks and Pocock
and Simon's procedure, but at the same time they yield two fewer failures
than the balanced designs. CARA 4 procedure has the power
of $80\%$ (same as for complete randomization), but it has, on average,
four fewer failures than the balanced designs. CARA 1 procedure
is the most skewed: the simulated allocation proportion for treatment
$A$ and the standard deviation is 0.40 (0.04), and it results, on
average, in six fewer treatment failures than in the balanced design
case. On the other hand, it is less powerful than balanced designs (the
average power is $76\%$).

%------------------
%t4 ###
\begin{table*}
\caption{Simulation results for
Model 3 with $\alpha_A-\alpha_B=-1.25$ and $n=160$\label{Table7-3}}
\begin{tabular*}{\textwidth}{@{\extracolsep{\fill}}l ccccc@{}}
\hline
\textbf{Procedure}
& $\bolds{\frac{N_A(n)}{n}}$ \textbf{(S.D.)}
& $\bolds{\frac{N_{A0}(n)}{N_0(n)}}$ \textbf{(S.D.)}
& $\bolds{d_{\mathrm{KS}}(z_2)}$ \textbf{(S.D.)}
& \textbf{Power}
& $\bolds{F(n)}$ \textbf{(S.D.)} \\
\hline
CRD & 0.50 (0.04) & 0.49 (0.05) & 0.14 (0.04) & 0.89 & 54 (6)\\
SPBD & 0.50 (0.01) & 0.50 (0.01) & 0.12 (0.03) & 0.89 & 54 (6) \\
P--S & 0.50 (0.01) & 0.50 (0.01) & 0.11 (0.03) & 0.90 & 54 (6) \\
CARA 1 & 0.39 (0.04) & 0.43 (0.04) & 0.13 (0.04) & 0.86 & 50 (6) \\
CARA 2 & 0.47 (0.03) & 0.48 (0.04) & 0.13 (0.04) & 0.90 & 53 (6) \\
CARA 3 & 0.48 (0.03) & 0.48 (0.04) & 0.13 (0.04) & 0.90 & 54 (6) \\
CARA 4 & 0.44 (0.03) & 0.45 (0.04) & 0.13 (0.04) & 0.89 & 51 (6) \\
CARA 5 & 0.47 (0.02) & 0.50 (0.03) & 0.12 (0.03) & 0.91 & 53 (5) \\
\hline
\end{tabular*}
\end{table*}
%------------------

The overall conclusion is that CARA procedures may be a good
alternative to covariate-adaptive procedures targeting balanced
allocations in the nonlinear response case. Although incorporating
responses in randomization induces additional variability of
allocation proportions, which may potentially reduce power, one can see
from our simulations that such an impact is not dramatic.

For CARA procedures, it is essential that the first allocations to
treatment groups are made by using some covariate-adaptive procedure
or the stratified block design, so that some data accrue and one can
estimate the unknown model parameters with reasonable accuracy. From
numerical experiments we have found that at least $80$ patients must be
randomized to treatment groups before m.l.e.'s can be computed.
Alternatively, one can check after each allocation the convergence of
the iteratively reweighted least squares \mbox{algorithm} for fitting the
logistic model, as Rosenberger,
Vidyashankar and Agarwal (\citeyear{rosVidAga2001}) did. However, due to the slow
convergence of m.l.e.'s, we have found that it is better, first, to
achieve reasonable quality estimators by using a covariate-adaptive
randomization procedure with good balancing properties (such as Pocock
and Simon's method).

From our simulations one can see that there are CARA procedures (such
as CARA 4 procedure) which have the same average power as complete
randomization, but at the same time they result in three to four fewer failures
than the balanced allocations. Such extra ethical savings together with
high power for showing treatment efficacy can be a good reason for
using CARA procedures to design efficient and more ethically attractive
clinical trials.

%s10 ###
\section{Discussion}\label{s10}

The design of clinical trials has become a rote exercise, often
driven by regulatory constraints. Boilerplate design sections in
protocols and grant proposals are routinely presented to steering
committees, review committees, and data and safety monitoring
boards. It is not uncommon for the randomization section of a
protocol to state ``double-blinded randomization will be performed''
with no further details. The fact that randomization is rarely if
ever used as a basis for inference means that the particular
randomization sequence is not relevant in the analysis, with the
exception that stratified designs typically lead to stratified
tests. Balance among important baseline covariates is seen to be an
essential cosmetic component of the clinical trial, and many
statisticians recommend adjusting for imbalanced covariates
following the trial, even if such analyses were not planned in the
design phase. While efficiency is usually gauged by a sample size
formula, the role that covariates play in efficiency, and the idea
that imbalances may sometimes lead to better efficiency and more
patients assigned to the superior treatment, are not generally
considered in the design phase of typical clinical trials.

In clinical trials with normally distributed outcomes, where it is assumed
that the variability of the outcomes is similar across treatments,
a balanced design across treatments and covariates will be the most
efficient. In these cases, if there are several important
covariates, stratification can be employed successfully, and if
there are many covariates deemed of sufficient importance,
covariate-adaptive randomization can be used to create balanced, and
therefore efficient, designs.

However, as we have seen, these simple ideas break down when there
are heterogeneous variances, including those found in commonly
performed trials with binary responses or survival responses. The
good news is that there are new randomization techniques that can be
incorporated in the design stage that can lead to more efficient and
more ethically attractive clinical trials. These randomization
techniques are based on the optimal design of experiments and also
tend to place more patients on the better treatment (Zhang et al., \citeyear{zhaHuCheChan2007}).
While more work needs to be done on the properties of
these procedures, we agree with Senn's comments that efficiency is
much more important than cosmetic balance.

The design of clinical trials is as important as the analysis of
clinical trials. Ethical considerations and efficiency should
dictate the randomization procedure used; careful selection of a
good design can save time, money, and in some cases patients' lives.
As Hu and Rosenberger (\citeyear{huRos2006}) point out, modern information
technology has progressed to the point where logistical difficulties
of implementing more complex randomization procedures are no longer
an issue. Careful design involves an understanding of both the
theoretical properties of a design in general, and simulated
properties under a variety of standard to worst-case models. In
some cases, the trade-offs in patient benefits and efficiency are so
modest compared to the relative gravity of the outcome, that
standard balanced designs may be acceptable. However, when outcomes
are grave, and balanced designs may produce severe inefficiency or
too many patients assigned to the inferior treatment, careful design
is essential.

\section*{Acknowledgments}

William F. Rosenberger is supported by
NSF Grant DMS-05-03718. The authors thank the referees for
helpful comments.

\vspace*{-1.5pt}
\end{document}